\documentclass[showpacs,reprint,prb,amsmath,amssymb,floatfix,superscriptaddress]{revtex4-1}
\usepackage{graphicx}											
\usepackage{amsmath}
\usepackage{amssymb}

\begin{document}

\title{Disorder-induced topological transitions in multichannel Majorana wires}

\author{B. Pekerten} 
\email{barisp@sabanciuniv.edu}
\affiliation{Faculty of Engineering and Natural Sciences, 
Sabanc{\i} University, Orhanl{\i} - Tuzla, 34956, Turkey}
\author{A. Teker} 
\affiliation{Faculty of Engineering and Natural Sciences, 
Sabanc{\i} University, Orhanl{\i} - Tuzla, 34956, Turkey}
\author{\"{O}. Bozat} 
\affiliation{Faculty of Engineering and Natural Sciences, 
Sabanc{\i} University, Orhanl{\i} - Tuzla, 34956, Turkey}
\author{M. Wimmer}
\affiliation{QuTech and Kavli Institute for Nanoscience, Delft University 
of Technology, Lorentzweg 1, 2628 CJ Delft, The Netherlands}
\author{\.{I}. Adagideli}
\email{adagideli@sabanciuniv.edu}
\affiliation{Faculty of Engineering and Natural Sciences, 
Sabanc{\i} University, Orhanl{\i} - Tuzla, 34956, Turkey}
\date{\today}

\begin{abstract}

In this work, we investigate the effect of disorder on the topological 
properties of multichannel superconductor nanowires. While the standard 
expectation is that the spectral gap is closed and opened at transitions that 
change the topological index of the wire, we show that the closing and opening 
of a \textit{transport} gap can also cause topological transitions, even in the 
presence of nonzero density of states across the transition. Such transport 
gaps induced by disorder can change the topological index, driving a 
topologically trivial wire into a nontrivial state or vice versa. We focus on 
the Rashba spin-orbit coupled semiconductor nanowires in proximity to a 
conventional superconductor, which is an experimentally relevant system, and 
obtain analytical formulas for topological transitions in these wires, valid 
for generic realizations of disorder. Full tight-binding simulations show 
excellent agreement with our analytical results without any fitting parameters.

\end{abstract}


% 74.78.Na 	Mesoscopic and nanoscale systems
% 74.45.+c 	Proximity effects; Andreev reflection; SN and SNS junctions
% 71.23.-k 	Electronic structure of disordered solids

\maketitle

\section{Introduction}

%%%%%%%%%%%%%%%%%%%%%%%%%%%%%%%%%%%%%%%%%%%%%%%%
%                                              %
% Introduction                                 %
%                                              %
%%%%%%%%%%%%%%%%%%%%%%%%%%%%%%%%%%%%%%%%%%%%%%%%

Topologically nontrivial phases are exotic states of matter that have an 
electronic band gap in their bulk and protected gapless excitations at their 
boundaries.~\cite{REF:Hasan10,REF:Qi11,REF:BookFranz13} Superconductors, being 
quasiparticle insulators, also feature topological phases with a quasiparticle 
gap in the bulk and excitations at their edges. For 1D systems, these edge 
states are fermionic zero-energy modes called Majorana 
states.~\cite{REF:Alicea12,REF:Leijnse12a,REF:Beenakker13,REF:BookBernevig13,REF:Elliott15} 
These states attracted intense attention owing to their non-Abelian nature, 
which led to proposals to use them as topological qubits immune to 
decoherence.~\cite{REF:Kitaev03,REF:Nayak08} Although predicted to appear in 
exotic condensed matter systems with unconventional superconducting 
pairing,~\cite{REF:Jackiw81,REF:Salomaa88,REF:Moore91,REF:Read00,REF:Ivanov01,REF:Kitaev01} 
recent proposals~\cite{REF:Alicea10,REF:Lutchyn10,REF:Sau10,REF:Oreg10} 
involving hybrid structures of more conventional materials have 
appeared.~\footnote{Note1} This led to the recent conductance measurements done 
on a proximity coupled InSb nanowire,~\cite{REF:Mourik12} which showed possible 
evidence of Majorana end states in the form of zero bias conductance peaks. 
Other experiments reported further observations of zero bias peaks in similar 
settings.~\cite{REF:Das12,REF:Deng12,REF:Finck13,REF:Churchill13,REF:Lee14} 
Very recently, scanning-tunneling spectroscopy experiments carried out on 
magnetic adatom chains on a conventional superconductor reported ZBPs at the 
ends of the chains.~\cite{REF:Nadj-Perge14} While it is compelling to interpret 
the observation of these ZBPs as signatures of Majorana states, the issue is 
still under intense discussion.~\footnote{Note2}

Semiconductor nanowire structures that are proximity-coupled to superconductors 
are technologically attractive platforms for Majorana physics. However, 
disorder has been prominently present in all such experimental samples to date. 
This led to a renewed interest in disordered superconducting wires, 
particularly focusing on the effects of disorder on Majorana 
states.~\cite{REF:Motrunich01,REF:Gruzberg05,REF:Akhmerov11,REF:Fulga11, 
REF:Potter11a,REF:Potter11b,REF:Stanescu11,REF:Brouwer11a,REF:Brouwer11b,REF:Sau12, 
REF:Lobos12,REF:Pientka13b,REF:DeGottardi13a,REF:Neven13,REF:Sau13, 
REF:Rieder13,REF:Chevallier13,REF:DeGottardi13b,REF:Jacquod13,REF:Adagideli14, 
REF:Hui14a} These works focused mostly on disordered \textit{p}-wave 
superconducting wires (PW wires) and showed that disorder is detrimental to the 
spectral gap as well as to the formation of Majorana fermions in both strictly 
1D systems~\cite{REF:Motrunich01,REF:Gruzberg05, 
REF:Brouwer11a,REF:Sau13,REF:Adagideli14,REF:Hui14a} and in multichannel 
wires.~\cite{REF:Stanescu11,REF:Pientka12,REF:Neven13,REF:Rieder13} In a recent 
study on the experimentally relevant hybrid structures with Rashba spin-orbit 
interaction (SOI) proximity coupled to an \textit{s}-wave superconductor (RSW 
nanowires for short), some of us showed that disorder need not be detrimental 
to and in fact can even \textit{create} topological order in strictly 1D 
wires.~\cite{REF:Adagideli14} We are not aware of a systematic study of the 
effects of disorder on the phase diagram of multichannel RSW nanowires.

In Majorana experiments, the subband spacing is typically considerably larger 
than the Zeeman splitting. For example, in InSb nanowires a subband spacing of 
order 15meV has been measured~\cite{REF:vanWeperen12,REF:Kammhuber16} together 
with a g-factor of 40 to 58. Zero bias peaks that might signal Majorana 
fermions in these works are typically measured at magnetic fields from 0.1mT - 
1T~\cite{REF:Mourik12,REF:Zhang16} and exceptionally up to 2.5T. In all of 
these cases the Zeeman splitting remains smaller than the level spacing. Hence, 
one can argue that RSW nanowires are more experimentally relevant than PW 
nanowires, which require Zeeman splitting be much larger than level spacing.

In this Manuscript, we investigate topological properties of disordered 
multichannel RSW and PW superconductor nanowires. The usual expectation for 
these nanowires is that if their topological state is switched by modifying 
certain external parameters (such as gate potential or magnetic field), the 
spectral gap will close and open concomitantly with this transition. We show 
that for disordered nanowires, the closing and opening of a \textit{transport} 
gap can cause further topological transitions, even in the presence of finite 
density of states (DOS), extending our earlier work on single channel 
wires~\cite{REF:Adagideli14} to multichannel wires. We derive analytical 
expressions for the boundaries of the topological phases of a disordered 
multichanneled RSW nanowire and find new topological regions in the phase 
diagram that show up as additional reentrant behavior in the experimentally 
relevant parameter regimes. In particular, new topological regions that show up 
in the low magnetic field limit, requires full description of all spin bands as 
shown by our analytical results (see 
Fig.~\ref{FIG:SWave_TB_Analytical_Numerical_Combined}). Hence, our results go 
beyond a simple \textit{p}-wave description that requires a fully spin 
polarized wire. Finally we perform numerical simulations using a tight-binding 
(TB) approach and find very good agreement with our analytical formulae.

This Manuscript is organized as follows: We begin the next section by 
specifying the system in question. We then derive the topological index in 
terms of the Lyapunov exponents and the effective superconducting length of the 
disordered multichannel RSW wire in 
subsection~\ref{SUBSECT:Topological_index_RSW}. In 
subsection~\ref{SUBSECT:Calculation_Topological_index_RSW}, we analytically 
calculate this topological index using experimentally relevant system and 
transport parameters and compare our results with numerical tight-binding 
simulations. We then present our conclusions, finding that in disordered 
multichannel RSW nanowires with experimentally relevant parameters, the 
topological phase diagram is fragmented and previously unreported reentrant 
topologically nontrivial regions appear. In the Appendices, we detail the 
calculation of the mean free path of the system 
(Appendix~\ref{SECT:Appendix_MFP_FermiGoldenRule}), detail our numerical 
simulations (Appendix~\ref{SECT:Appendix_TB}), present a full bandwith versions 
of our plots in the main text as opposed to the low energy region 
(Appendix~\ref{SECT:Appendix_FullBWPlots}), and finally present our plots 
similar to the RSW system but preoduced for a \textit{p}-wave nanowire with 
disorder, as system previously studied in literature, for completeness and 
comparison (Appendix~\ref{SECT:Appendix_PWave}).

\section{Topological order in disordered multichannel wires}\label{SECT:DisorderedTSWires}

%%%%%%%%%%%%%%%%%%%%%%%%%%%%%%%%%%%%%%%%%%%%%%%%
%                                              %
% Topological order in Disordered              %
%            quasi-1D Wires                    %
%                                              %
%%%%%%%%%%%%%%%%%%%%%%%%%%%%%%%%%%%%%%%%%%%%%%%%

\begin{figure}
	\centering \includegraphics[width=0.9\columnwidth]{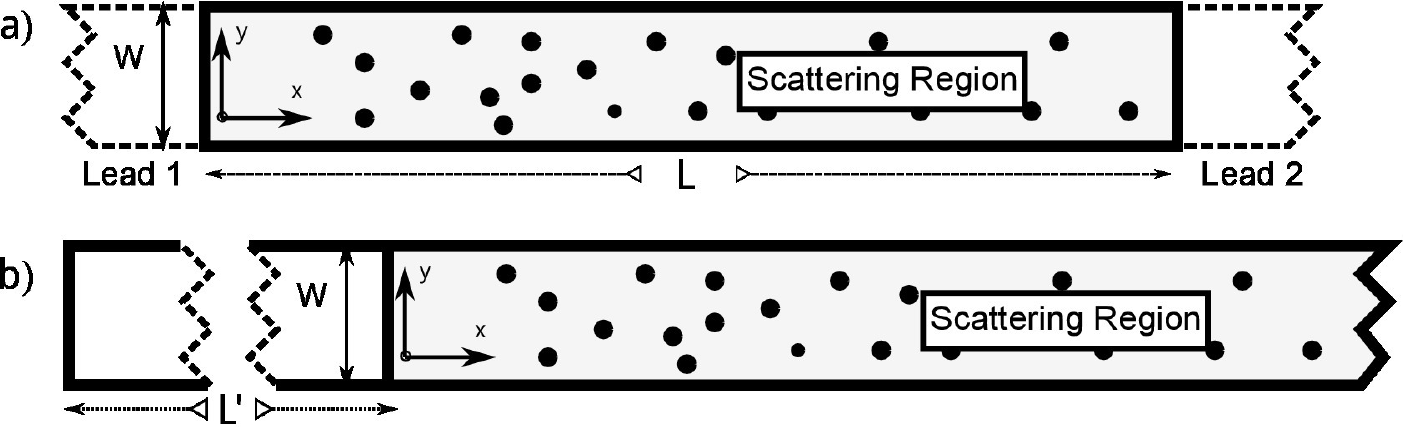} 
	\caption{The multichanneled nanowire of width $W$, which is an RSW 
	topological superconductor with a Gaussian disorder having an average value 
	$\left\langle V\right\rangle=0$. a) In the leads, we take 
	$\alpha_\textrm{SO}$, $\Delta$ and $V(x,y)$ to be zero, making the leads 
	metallic. Our analytical results assume a semi-infinite wire 
	($L\rightarrow\infty$), whereas in our numerical full tight-binding 
	calculations we use wires of length $L\gg l_\text{MFP}, \xi, 
	l_\textrm{SO}$. b) The form of the wire used to construct the Majorana 
	solutions in section \ref{SUBSECT:Topological_index_RSW}. The part of the 
	wire left of the scattering region is again metallic.}\label{FIG:2DWire}
\end{figure}

In this section, we investigate the topological properties of multichanneled 
topological superconductor nanowires. Such wires are experimentally realized 
by proximity coupling a semiconductor nanowire 
with Rashba spin-orbit interaction to an \textit{s}-wave superconductor (RSW, 
see Fig.~\ref{FIG:2DWire} (a)). The quasiparticles in RSW nanowires are described 
by the following Bogoliubov--de Gennes (BdG) 
Hamiltonian:~\cite{REF:Lutchyn10,REF:Oreg10,REF:deGennes99}
\begin{align}\label{EQN:SWaveHamiltonian}
H &= \int \Psi^\dagger \, \mathcal{H}_\mathrm{BdG} \, \Psi \, d\mathbf{r} \nonumber\\
\mathcal{H}_\mathrm{BdG} &=\left(h_0 +\alpha_\textrm{SO}(\mathbf{p}\times\mathbf{\sigma})\right)\tau_z+B\sigma_x+\Delta\tau_x,
\end{align}
where $h_0 = \varepsilon(p) + V(\mathbf{r})$, $\Psi^\dagger = 
[\psi_\uparrow^\dagger, \psi_\downarrow^\dagger, \psi_\downarrow, 
-\psi_\uparrow]$ is the Nambu spinor with $\psi_{\uparrow(\downarrow)}$ being 
the destruction operator for an electron with spin up(down). The kinetic energy 
term $\varepsilon(p)$ is given by $\frac{\mathbf{p}^2}{2m}-\mu$ in a continuum 
system. We consider a 2D wire with $\mathbf{p}=(p_x,p_y)$.  The on-site 
potential is given by $V(\mathbf{r})$, $\mu$ is the chemical potential measured 
from the bottom of the band, $\alpha_\textrm{SO}$ is the spin-orbit coupling 
(SOC) strength, $B$ is the Zeeman field and $\Delta$ is the proximity-induced 
\textit{s}-wave superconducting gap. The Pauli matrices $\sigma_i$ ($\tau_i$) 
act on the spin (electron-hole) space. 

In the limit of large $B$, the wire is completely spin polarized. Then the 
low-energy quasiparticles are described by an effective \textit{p}-wave 
Hamiltonian as discussed in previous literature.~\cite{REF:Adagideli14, 
REF:Akhmerov11,REF:Brouwer11b,REF:DeGottardi13a,REF:Fulga11,REF:Hui14a, 
REF:Lobos12,REF:Rieder13,REF:Potter11a,REF:Potter11b,REF:Rieder12,REF:Sau12, 
REF:Sau13} For completeness, we discuss this limit in 
Appendix~\ref{SECT:Appendix_PWave}.

The Hamiltonian in Eq.~(\ref{EQN:SWaveHamiltonian}) is in the 
Altland-Zirnbauer (AZ) symmetry class D (\textit{class D} for short) in 
two dimensions~\cite{REF:Altland97} with a topological number 
$Q_\textrm{D} \in \mathbb{Z}_2$ . In the absence of SOC along the 
$y$-direction, i.e. when the $\alpha_\textrm{SO} \,p_y \,\sigma_x 
\tau_z$ term is set to zero, this Hamiltonian also possesses a chiral 
symmetry, placing it into AZ symmetry class BDI (\textit{class BDI} for 
short) with an integer topological number $Q_\textrm{BDI}\in 
\mathbb{Z}$.~\cite{REF:Tewari12,REF:Rieder13} In the thin wire limit, 
i.e. $W\ll l_\textrm{SO}$, chiral symmetry breaking terms are 
$\mathcal{O}\left(( W/l_\textrm{SO})^2\right)$. Hence, the system in 
Eq.~(\ref{EQN:SWaveHamiltonian}) has an approximate chiral 
symmetry.~\cite{REF:Rieder12,REF:Tewari12,REF:Diez12} We show in the 
next section that the class-BDI (chiral) topological number 
$Q_\textrm{BDI}\in \mathbb{Z}$ and the class-D topological number are 
related as $Q_\textrm{D} = (-1)^{Q_\textrm{BDI}}$ (see 
Eq.~(\ref{EQN:Q_D})).~\cite{REF:Fulga11}

\subsection{Topological index for a disordered multichannel \textit{s}-wave wire}\label{SUBSECT:Topological_index_RSW}

%%%%%%%%%%%%%%%%%%%%%%%%%%%%%%%%%%%%%%%%%%%%%%%%
%                                              %
% Subsection: Topological index for a          %
%             Disordered RSW                   %
%                                              %
%%%%%%%%%%%%%%%%%%%%%%%%%%%%%%%%%%%%%%%%%%%%%%%%

To obtain the relevant topological index that counts the number of the Majorana 
end states for a RSW wire with disorder, we start with the BdG Hamiltonian 
$\mathcal{H}_\mathrm{BdG}$ in Eq.~(\ref{EQN:SWaveHamiltonian}). Following 
Adagideli \textit{et al.},~\cite{REF:Adagideli14} we perform the unitary 
transformation $\mathcal{H}_\mathrm{BdG}\rightarrow 
\mathcal{H}_\mathrm{BdG}'=\mathcal{U}^\dagger 
\mathcal{H}_\mathrm{BdG}\mathcal{U} $, where 
$\mathcal{U}=(1+i\sigma_x)(1+i\tau_x) \left[1+\sigma_z+(1-\sigma_z) 
\tau_x\right]/4$. Having thus rotated the Hamiltonian to the basis that 
off-diagonalizes its dominant part and leaves the small chiral symmetry 
breaking terms $\tau_z\sigma_z$ in the diagonal block, we obtain
\begin{align}\label{EQN:OffDiagonalizedHamiltonian}
\mathcal{H}_\mathrm{BdG}'	&= -\tau_y \left(\sigma_z\,h_0+\alpha_\textrm{SO}\,p_x\right) + \tau_x \left(B\,\sigma_x+\Delta\right)\nonumber\\
							&\quad +\,\tau_z\sigma_y\,\alpha_\textrm{SO}\,p_y\,.
\end{align}

We first set the chiral symmetry breaking term $\tau_z\,\sigma_y\,\alpha_\textrm{SO}\, 
p_y$ to zero and focus on $E=0$. The eigenvalue equation then decouples into 
the upper and lower spinor components. The solutions are of the form 
$\chi_+=(\phi_+,0)^T$ and $\chi_-=(0, \phi_-)^T$ where $\phi_\pm$ obey the 
following equation:
\begin{align}\label{EQN:Appendix:NormalStateHamiltonian}
\left(\varepsilon(p) \sigma_z - i\, p_x \alpha_\textrm{SO} \sigma_x \mp B \mp \Delta \sigma_x \right)\, \phi_\pm &= 0.
\end{align}
Here, we have performed an additional rotation $\sigma_z \rightarrow \sigma_y$, 
$\sigma_y \rightarrow -\sigma_z$ and premultiplied with $\pm \sigma_x$. We note 
that the operator acting on $\phi_\pm$ is not Hermitian.

We now perform a gauge transformation $\phi_\pm(x,y) \rightarrow 
e^{-\kappa_\alpha x} \phi_\pm(x, y)$ with a purely imaginary parameter 
$i\kappa_\alpha$. We take $\kappa_\alpha$ to be of first order in 
$\alpha_\textrm{SO}$ and identify the following terms in the nonhermitian 
operator in Eq.~(\ref{EQN:Appendix:NormalStateHamiltonian}) in order of 
increasing power of $\alpha_\textrm{SO}$:
\begin{align}\label{EQN:ExpandForSmallAlpha}
H_0	&= h_0(p; x,y)\sigma_z \mp B \mp \Delta \sigma_x \nonumber\\
H_1	&= \frac{i \hbar \kappa_\alpha p_x}{m} \sigma_z - i \alpha_\textrm{SO} p_x \sigma_x \nonumber\\
H_2	&= -\frac{\hbar^2 \kappa_\alpha^2}{2m} \sigma_z + \hbar \alpha_\textrm{SO} \kappa_\alpha \sigma_x, 
\end{align}
where we have indicated the $(x,y)$ dependence of $h_0(p; x,y)$ through the 
potential $V(x,y)$. We absorb $H_2$ into $H_0$ by redefining $\mu$ and 
$\Delta$. We now identify $\kappa_\alpha$ with the inverse of the effective 
superconducting length $\xi_\textrm{eff}$, setting $\kappa_\alpha = 
\mp \xi_\textrm{eff}^{-1} = \mp m \alpha_\textrm{SO} \Delta/\hbar \epsilon$ with 
$\epsilon = \sqrt{B^2-\Delta^2}$. With this choice, $\{H_0,H_1\}_+=0$, which 
allows us to write the local solutions as follows:
\begin{align}\label{EQN:GeneralEigensolution} 
\phi_\pm	&= \sum_n\xi_\pm(\epsilon) {\rm e}^{\pm \kappa_\alpha x} \big(A_n f_n(x,y;\epsilon) + B_n g_n(x,y;\epsilon) \big) \nonumber \\ 
			&\quad+ \xi_\pm(-\epsilon) {\rm e}^{\mp \kappa_\alpha x} \big( C_n f_n(x,y;-\epsilon) + D_n g_n(x,y;-\epsilon) \big), 
\end{align} 
where $\xi_\pm(\epsilon)$ are the eigenvectors  of the $2\times 2$ matrix 
$\epsilon \sigma_z \mp \Delta \sigma_x$ with eigenvalue $\pm |B|$ and $f_n$ and 
$g_n$ are the local solutions of the equation $h_0 \psi=\epsilon \psi$. The 
presence of a multiple number of local solutions, which is the new aspect of 
the present problem, reflects the multichannel nature of the wire.

We then consider a semi-infinite wire ($x>0$, $0<y<W$) described by the Hamiltonian in 
Eq.~(\ref{EQN:SWaveHamiltonian}) with Gaussian disorder. After going through the 
steps described above, we choose without loss of generality $f_n$ to be the 
decaying and $g_n$ the increasing function of $x$. We invoke a well known 
result from disordered multichannel normal state wires and express the 
asymptotic solutions as $f_n = e^{-\Lambda_n x} u_n(x,y)$ and $g_n = 
e^{\Lambda_n x} v_n(x,y)$ where $u_n(x,y), v_n(x,y)$ are ${\cal O}(1)$ 
functions as $x\rightarrow \infty$ and $\Lambda_n>0$ are the \textit{Lyapunov 
exponents}.~\cite{REF:Beenakker97,REF:Fulga11,REF:DeGottardi13a,REF:Rieder13, 
REF:Adagideli14} 

We now focus on a tight-binding system, where the number of Lyapunov exponents 
$N_\textrm{max}$ is finite. (In the continuum limit, we have 
$N_\textrm{max}\rightarrow \infty$.)  For the boundary conditions at $x=0$, we 
first extend the hardwall back to $x=-L'$ with $L'$ a small value, and consider 
a normal metal in the strip $-L'<x<0$ and $0<y<W$ (see Figure \ref{FIG:2DWire} 
(b); in Eq.~(\ref{EQN:SWaveHamiltonian}), $\alpha_\textrm{SO}=0$, $\Delta=0$, 
$V(x,y)=0$).  The hardwall boundary condition at $x=-L'$ can be expressed as 
$\underline{\underline{R}} \cdot \underline{b}_+ = \underline{b}_-$ with 
$\underline{b}_+ \equiv (\ldots, A_n, C_n, \ldots)^T$, $\underline{b}_- \equiv 
(\ldots, B_n, D_n, \ldots)^T$ and $\underline{\underline{R}}$ as the extended 
reflection matrix.~\cite{REF:Mello04} We therefore have $2N_\textrm{max}$ 
boundary conditions, leaving $2N_\textrm{max}$ of the $4N_\textrm{max}$ 
parameters undetermined.

The boundary conditions at $x\rightarrow\infty$ require that $\phi_\pm$ have 
only exponentially decaying solutions. We focus on the $B>\Delta$ case, 
yielding real $\kappa_\alpha$ and $\epsilon$. (As discussed in References 
\cite{REF:Sau10} and \cite{REF:Oreg10}, the $B<\Delta$ case yields no 
solutions.) We take $\kappa_\alpha>0$ for definiteness. (The following 
arguments can be extended trivially to the $\kappa_\alpha<0$ case.) The 
exponential asymptotic factors in the solutions contain a factor of $e^{\pm 
\kappa_\alpha x}$ in various sign combinations, affecting the overall 
convergence at $x\rightarrow\infty$. In particular, the solutions $\phi_+$ have 
exponential factors of $e^{(\kappa_\alpha-\lambda_n(\epsilon))x}$, 
$e^{(\kappa_\alpha+\lambda_n(\epsilon))x}$, 
$e^{(-\kappa_\alpha-\lambda_n(-\epsilon))x}$ and 
$e^{(-\kappa_\alpha+\lambda_n(-\epsilon))x}$, whereas the $\phi_-$ solutions 
have the same form of exponential factors with the sign of $\kappa_\alpha$ 
switched. For $|\kappa_\alpha|$ smaller than all Lyaponov exponents, all $B_n$ 
and $D_n$ are set to zero as they would represent diverging solutions at $x 
\rightarrow \infty$. There are therefore $2N_\textrm{max}$ more conditions , 
bringing the total up to $4N_\textrm{max}$, to determine a total of 
$4N_\textrm{max}$ parameters, yielding only accidental solutions. However, for 
a given $n=n_*$, if $\textrm{min}\,(\lambda_{n_*}(\epsilon), 
\lambda_{n_*}(-\epsilon)) < \kappa_\alpha < 
\textrm{max}\,(\lambda_{n_*}(\epsilon), \lambda_{n_*}(-\epsilon))$, there are 
three growing solutions for one of the $\phi_\pm$ sectors and only one for the 
other sector. (If $\lambda_{n_*}(\epsilon)<\lambda_{n_*}(-\epsilon)$, the 
$\phi_+$ sector has the three growing solutions and vice versa.) The sector 
with three growing solutions thus has the number of boundary conditions 
increased by one and the other sector has the number of boundary conditions 
decreased by one. If any sector has more than $4N_\textrm{max}$ boundary 
conditions in total, then there are no solutions for that sector. Therefore, 
the BDI topological number $Q_\textrm{BDI}\in \mathbb{Z}$ is given by the 
number of free parameters, which is equal to $4N_\textrm{max}$ minus the total 
number of equations arising from the boundary condition at $x=-L'$. We obtain:
\begin{align}\label{EQN:Q_BDI} 
Q_\textrm{BDI} &= \sum_{n}\Theta\left(\xi_\text{eff}^{-1} - \Lambda_n(\epsilon) \right)\, \Theta\left(\Lambda_n(-\epsilon) - \xi_\text{eff}^{-1} \right) \nonumber\\
				&\quad -\sum_{n}\Theta\left(\xi_\text{eff}^{-1} - \Lambda_n(-\epsilon) \right)\, \Theta\left(\Lambda_n(\epsilon) - \xi_\text{eff}^{-1} \right).
\end{align}
We see that each Lyapunov exponent pair $\Lambda_n(\pm\epsilon)$ contributes a 
topological charge $Q^{(n)}_\textrm{BDI}$ to the overall topological charge. 
Hence $Q_\textrm{BDI}=\sum_n Q^{(n)}_\textrm{BDI}$, where
\[ 
Q^{(n)}_\textrm{BDI} =
\begin{cases}
+1 \quad \textrm{if } \Lambda_n(-\epsilon) > \xi_\textrm{eff}^{-1} > \Lambda_n(\epsilon) \\
-1 \quad \textrm{if } \Lambda_n(-\epsilon) < \xi_\textrm{eff}^{-1} < \Lambda_n(\epsilon) \\
{\phantom +}0 \quad \textrm{otherwise.}
\end{cases}
\]
We thus generalize the resuls of Ref.~\cite{REF:Adagideli14} to a multichannel 
RSW wire. We note, however, that the total number of Majorana end states for a 
multichannel RSW wire in class BDI, given by $|Q_\textrm{BDI}|$, is not equal 
to sum of the Majorana states per Lyapunov exponent pair, 
i.e.~$|Q_\textrm{BDI}| \ne \sum_n |Q^{(n)}_\textrm{BDI}|$.

We now consider the full Hamiltonian in Eq.~(\ref{EQN:SWaveHamiltonian}) with 
the chiral symmetry breaking term included. This Hamiltonian in two dimensions 
is in class D and only approximately in class BDI. The chiral symmetry breaking 
term pairwise hybridizes the Majorana states described above, moving them away 
from zero energy. However, because of the particle-hole symmetry in the 
topological superconductor, any disturbance or any perturbation that is higher 
order in $\alpha_\textrm{SO}$ can only move the solutions away from zero energy 
eigenvalue in pairs; i.e. for any solution moving away from zero eigenvalue 
towards a positive value, a matching solution must move to a negative 
eigenvalue. Therefore, the number of zero eigenvalue solutions changes in 
pairs. Hence, the parity doesn't change. The parity changes, however, every 
time one of the Lyapunov exponents passes through the value of 
$\xi_\text{eff}^{-1}$. We therefore arrive at the class D topological index 
$Q_\textrm{D} = (-1)^{Q_\textrm{BDI}}$ as~\cite{REF:Fulga11}
\begin{align}\label{EQN:Q_D}
Q_\textrm{D}	&= \prod_{n,\pm}{\rm sgn}\big(\Lambda_n(\pm\epsilon)\,\xi_\text{eff} -1\big),
\end{align}
indicating that there's a class D Majorana solution at zero energy 
($Q_\textrm{D} = -1$) if there are an odd number of BDI Majorana states per 
edge. Therefore, for the topological state of the RSW wire to change from 
trivial to nontrivial or vice versa, it is necessary and sufficient to have 
$Q_\textrm{BDI}$ described in Eq.~(\ref{EQN:Q_BDI}) change by one. The above 
equation thus constitutes the multichannel generalization of Eq.(7) of 
Ref.~\cite{REF:Adagideli14}.

To calculate the topological index $Q_\textrm{D}$ in Eq.~(\ref{EQN:Q_D}), we relate 
the Lyapunov exponents in Eq.~(\ref{EQN:Q_BDI}) to transport properties, namely 
the mean free path, of a disordered wire. We first note that as $L \rightarrow 
\infty$, the Lyapunov exponents $\Lambda_n$ are self-averaging, with a mean 
value $\bar{\Lambda}_n$ given by
\begin{align}\label{EQN:LyapunovExponent_from_MFP}
\bar{\Lambda}_{n}(\mu_\textrm{eff})&=\frac{n}{(\bar{N}(\mu_\textrm{eff})+1)\,l_\text{MFP}}
\end{align}
where $\mu_\textrm{eff}=\mu\pm\epsilon$, $\bar{N}(\mu_\textrm{eff}) = \lfloor W 
k_F(\mu_\textrm{eff})/\pi \rfloor$, $k_F=\sqrt{2m\mu_\textrm{eff}/\hbar^2}$, 
$n\in1\ldots \bar{N}(\mu_\textrm{eff})$ and $l_\text{MFP}$ is the MFP of the 
disordered wire.~\cite{REF:Beenakker97} We use Fermi's Golden Rule to 
approximate the mean free path $l_\textrm{MFP}$ by calculating the lifetime of 
a momentum state and multiplying it with the Fermi speed. We obtain, for a 
quadratic dispersion relation $\varepsilon(p) = p^2/2m-\mu$, 
\begin{align}\label{EQN:inverseMFP_FreeElectron_EndResult}
l_\text{MFP}^{-1}&=\frac{4m^2\gamma}{\hbar^4\pi k_F}\,\zeta_N^{-1},
\end{align}
where $\zeta_N^{-1}$ is a dimensionless number whose detailed form is given in 
Eq.~(\ref{EQN:Appendix:inverseZetaN_FreeElectron}). The details of this 
calculation can be found in Appendix~\ref{SECT:Appendix_MFP_FermiGoldenRule}. 

In order to compare our numerical tight-binding results with the analytical 
results obtained through Eq.~(\ref{EQN:Q_D}) and (\ref{EQN:Q_BDI}), we also 
calculate the mean free path $l_{\text{MFP}}^\text{TB}$ for a tight-binding 
(TB) dispersion relation $\varepsilon(k_{x,n}) = 2t \, 
\left(2-\cos{(k_{x,n}a)}-\cos{(n\pi a/W)}\right)$, where $t$ is the hopping 
parameter, $a$ is the lattice parameter for the TB lattice, $W$ is the width of 
the lattice and $k_{x,n}$ is defined through $k_{x,n}^2+k_{y,n}^2=k_F^2$ with 
$k_{y,n}=n\pi/W$. We obtain
\begin{align}\label{EQN:inverseMFP_TB_EndResult}
(l_{\text{MFP}}^\text{TB})^{-1}&=\frac{\gamma}{\bar{N}^\textrm{TB} Wa^2t^2}\,(\zeta_N^{\text{TB}})^{-1}.
\end{align}
where $\bar{N}^\textrm{TB}$ is given by $\lfloor (W/\pi a) 
\arccos{(1-\varepsilon/2t)} \rfloor$ for $0<\varepsilon <4t$ and $\lfloor 
(W/\pi a) \arccos{(1-(4-\varepsilon/2t))} \rfloor$ for $4t<\varepsilon <8t$. 
The details of the calculation and the dimensionless constant 
$\zeta_N^{\text{TB}}$ are again found in 
Appendix~\ref{SECT:Appendix_MFP_FermiGoldenRule}.

The topological phase boundaries, shown in 
Figures~\ref{FIG:SWave_TB_NumericalAnalytical_MuvsB_Zoomed} and 
\ref{FIG:SWave_TB_Analytical_Numerical_Combined} as the bold black lines, are 
calculated by equating $\xi^{-1}$ to $\Lambda_n$ obtained from 
Eq.~(\ref{EQN:LyapunovExponent_from_MFP}) and 
(\ref{EQN:inverseMFP_TB_EndResult}). We thus obtain the critical field $B^*$ at 
which the system goes through a topological phase transition via thie following 
implicit equation:
\begin{align}\label{EQN:Topological_phase_boundaries_TB}
B^* &= \Delta\,\sqrt{\beta \, \Gamma_n^\textrm{TB}\left(\mu_\textrm{eff}(B^*)\right) +1}
\end{align}
where $\beta = (Wa^2t^2/\gamma l_\textrm{SO})^2$, $\mu_\textrm{eff}(B^*) = \mu \pm 
\sqrt{(B^*)^2+\Delta^2}$ and 
\begin{align*}
\Gamma_n^\textrm{TB}\left(\mu_\textrm{eff}\right) 	&= \left(\frac{\bar{N}^\textrm{TB}\left(\mu_\textrm{eff}\right)}{n}\right)^2\nonumber\\
														&\qquad \times\left(\zeta_N^{\text{TB}}\left(\mu_\textrm{eff}\right)\right)^2 \left(\bar{N}^\textrm{TB}\left(\mu_\textrm{eff}\right)+1)\right)^2. 
\end{align*}
Equation~(\ref{EQN:Topological_phase_boundaries_TB}) constitutes the central 
finding of our paper. It is an analytical expression that determines all
topological phase boundaries of a multichannel disordered wire.

An experimentally interesting point is the largest values of various system 
parameters that allow a topological transition. Using 
Equations~(\ref{EQN:Q_BDI}) and (\ref{EQN:Q_D}), we estimate the upper critical 
field $B^*|_\gamma\,$, i.e. the minimum value of $B$ above which the system is 
always in a topologically trivial state at a given disorder strength $\gamma$, 
as
\begin{align}\label{EQN:Bmax} 
B^*|_\gamma &\sim \Delta \, \frac{l_\textrm{tr}^\textrm{max}}{l_\textrm{SO}},
\end{align} 
where $l_\textrm{tr}^\textrm{max} = \textrm{max}(\{\Lambda_n^{-1}\})$ is the 
maximum localization length achievable in the system. For a fixed nonzero 
disorder, $B^*|_{\gamma> 0}$ is infinite for a continuum system as the 
localization length increases indefinitely with increasing Fermi energy. For a 
TB system, the upper critical field $B^*|_{\gamma> 0}$ is finite because the 
localization length is bounded in TB systems. For a clean wire, 
$B^*|_{\gamma=0}$ is infinite for both the TB and the continuum models.

\subsection{Numerical simulations}\label{SUBSECT:Calculation_Topological_index_RSW}

%%%%%%%%%%%%%%%%%%%%%%%%%%%%%%%%%%%%%%%%%%%%%%%%
%                                              %
% Subsection: Numerical simulations            %
%                                              %
%%%%%%%%%%%%%%%%%%%%%%%%%%%%%%%%%%%%%%%%%%%%%%%%

\begin{figure}
	%\centering 
	\hspace*{-1cm}
	\includegraphics[width = 0.98\columnwidth]{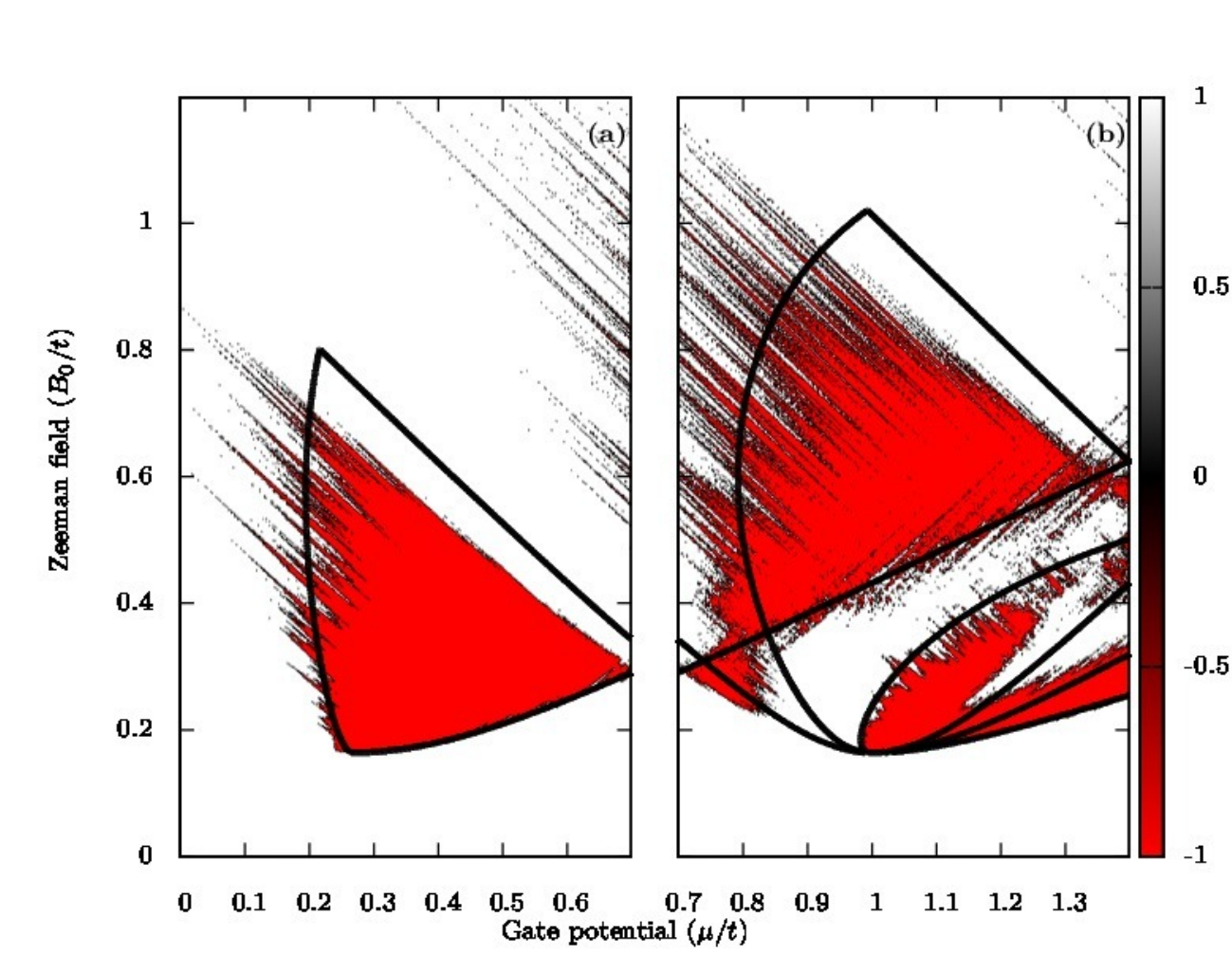}	
	\caption{(Color online) $\mu$ vs. $B$ vs. $Q_\textrm{D}$ for a five-channel 
		system (compare with 
		Figs.~\ref{FIG:Appendix:SWave_TB_NumericalAnalytical_MuvsB_FullBandwidth} 
		and \ref{FIG:Appendix:SWave_TB_Analytical_MuvsB_FourChannels}.) The 
		background red-white colors are obtained using a numerical 
		tight-binding simulation with $L=30000a$ and $W=5a$, while the black 
		lines, which represent the topological phase boundaries, are obtained 
		analytically using Eq.~(\ref{EQN:Q_D}). Here, $V_0=\sqrt{\gamma/ 
		a^2}=0.2 t$, $\alpha_\textrm{SO}=0.02\hbar/ma$ ($l_\textrm{SO}=4.08\mu 
		m$) and $\Delta=0.164t$, where $t=\hbar^2/2ma^2$ and 	$a=0.01 
		l_\textrm{SO}$ is the tight-binding lattice spacing. The fragmented 
		nature of the topological phase diagram  seen in (b) cannot be 
		explained in a \textit{p}-wave picture. See 
		Appendix~\ref{SECT:Appendix_TB} for a discussion of corresponding 
		experimental parameters.} 
	\label{FIG:SWave_TB_NumericalAnalytical_MuvsB_Zoomed}
\end{figure}

%%%%%%%%%%%%%%%%%%%%%%%%%%%%%%%%%%%%%%%%%%%%%%%%%%%%%%%%%%%%%%%%%%%%%%%%%%%%%%%%%%%%%%%%%%%%%%%%%%%%%%%%%%%%%%%%%%%%%%%%%%%%%%%%%%%%%%%%%%%

\begin{figure}
	\hspace*{-1cm}
	\includegraphics[width=1.0\columnwidth]{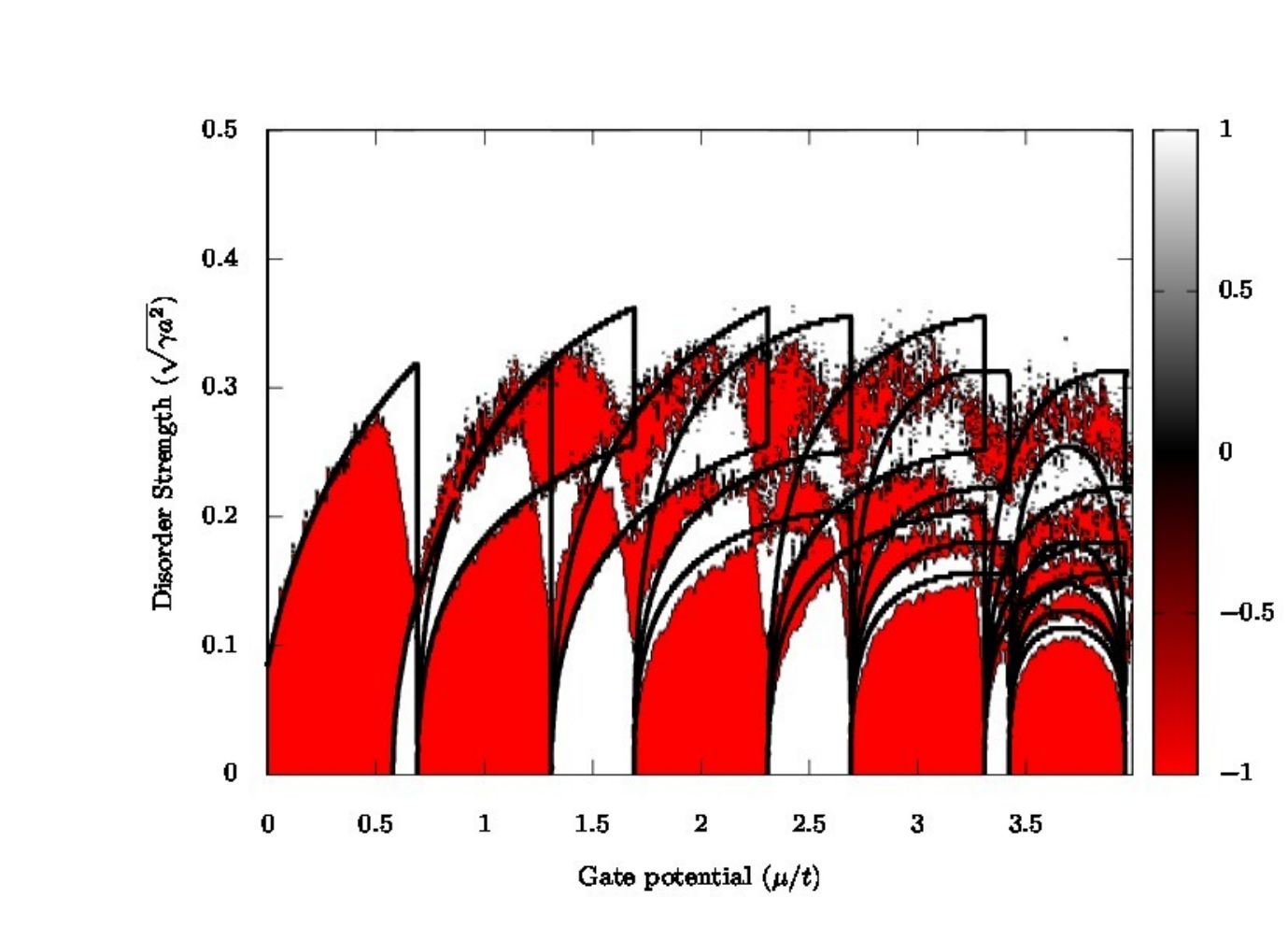}
	\caption{(Color online) 
		$\mu$ vs. $V_0 = \sqrt{\gamma/a^2}$ vs. $Q$ for a multichannel
		RSW wire. The black lines, which represent topological phase boundaries, are 
		obtained analytically using Eq.~(\ref{EQN:Q_D}). The background 
		red-white colors are obtained using tight-binding numerical simulations 
		with $L=60000a$. In both cases, $W=4a$, 
		$\alpha_\textrm{SO} =0.015\hbar/ma$, $\Delta = 0.20t$ and 
		$B=0.35t$, where $t=\hbar^2/2ma^2$ is the tight-binding hopping parameter and 
		$a$ is the TB lattice spacing. See Appendix~\ref{SECT:Appendix_TB} for 
		a discussion of corresponding experimental parameters.}		
\label{FIG:SWave_TB_Analytical_Numerical_Combined}
\end{figure}

%%%%%%%%%%%%%%%%%%%%%%%%%%%%%%%%%%%%%%%%%%%%%%%%%%%%%%%%%%%%%%%%%%%%%%%%%%%%%%%%%%%%%%%%%%%%%%%%%%%%%%%%%%%%%%%%%%%%%%%%%%%%%%%%%%%%%%%%%%%

In this section, we obtain the topological index of a disordered multichannel 
wire numerically and compare it with our analytical results from the previous 
section. For our numerical simulations, we take the TB form of the Hamiltonian 
in Eq.~(\ref{EQN:SWaveHamiltonian}) whose details can be found in the 
Appendix~\ref{SECT:Appendix_TB}. We consider a wire of length $L\gg 
l_\text{MFP}$, $\xi$ or $\l_\textrm{SO}$, with metallic leads 
($\alpha_\textrm{SO}=0$, $\Delta=0$ and $V(x,y)=0$ in the leads). We use the 
results of Fulga \textit{et al.} to obtain the topological quantum number of 
the disordered multichannel wire from the scattering matrices of the 
wires.~\cite{REF:Fulga11} For a semi-infinite wire in the symmetry class D, the 
topological charge is given by $Q_\textrm{D} = \text{det}(r)$ where $r$ is the 
reflection matrix. For a quasiparticle insulator, this determinant can only 
take the values $\pm 1$. However, for a finite system  this determinant can in 
general have any value in the $[-1,1]$ interval. We obtain the reflection 
matrix of the TB system in our numerical TB simulations using the Kwant 
library~\cite{REF:Kwant14} and then use this relation to calculate 
$Q_\textrm{D}$. We plot the topological phase diagram in 
Figures~\ref{FIG:SWave_TB_NumericalAnalytical_MuvsB_Zoomed} and 
\ref{FIG:SWave_TB_Analytical_Numerical_Combined}, where the red and white 
colors represent $Q_\textrm{D}=-1$ and $Q_\textrm{D}=+1$ respectively. 

Figure~\ref{FIG:SWave_TB_NumericalAnalytical_MuvsB_Zoomed} exemplifies our 
central result given in Eq.~(\ref{EQN:Topological_phase_boundaries_TB}). We find 
that for a nearly depleted wire 
(Fig.~\ref{FIG:SWave_TB_NumericalAnalytical_MuvsB_Zoomed}a), the topological 
phase merely shifts to the higher values of the chemical potential in agreement 
with Ref.~\cite{REF:Adagideli14}. For higher chemical potentials/doping, we 
observe a fragmented topological phase diagram 
(Fig.~\ref{FIG:SWave_TB_NumericalAnalytical_MuvsB_Zoomed}b). We find good 
agreement with our analytical results from 
Eq.~(\ref{EQN:Topological_phase_boundaries_TB}). We note in passing that, this 
fragmentation cannot be explained by a simple \textit{p}-wave picture as these 
topological phases arise despite the incomplete spin-polarization of the wire 
under a low magnetic field. For a full phase diagram over the entire bandwidth, 
but for slightly different material parameters, see 
Figure~\ref{FIG:Appendix:SWave_TB_NumericalAnalytical_MuvsB_FullBandwidth}, 
where the reentrant phases are apparent.

In Fig.~\ref{FIG:SWave_TB_Analytical_Numerical_Combined}, we plot the 
topological number $Q_\textrm{D}$ as a function of $\mu$ and the disorder 
strength $\sqrt{\gamma/a^2}$ for a constant $B_\textrm{Zeeman}$ over the full 
TB bandwidth. The reentrant nature of the topological phase diagram can also be 
seen in this plot, for example, by following the $\mu=1.5$ line as $\gamma$ is 
increased. As the disorder strength increases, series of topological 
transitions occur, similar to the PW wire.~\cite{REF:Rieder13} However, unlike 
the PW wire, the number of transitions is given by 
$\bar{N}(\mu+\epsilon)+\bar{N}(\mu-\epsilon)$ rather than $\bar{N}(\mu)$, with 
$\bar{N}(\mu)$ defined as $\bar{N}(\mu_\textrm{eff}) = \lfloor W 
k_F(\mu_\textrm{eff})/\pi \rfloor$. For further discussion of the emergence of 
effective p-wave picture at high magnetic fields, see 
Appendix~\ref{SECT:Appendix_FullBWPlots}.

\section{Conclusion}

%%%%%%%%%%%%%%%%%%%%%%%%%%%%%%%%%%%%%%%%%%%%%%%%
%                                              %
% Conclusion                                   %
%                                              %
%%%%%%%%%%%%%%%%%%%%%%%%%%%%%%%%%%%%%%%%%%%%%%%%

In summary, we investigate the effect of disorder in multichannel Rashba SOC 
proximity-induced topological superconductor nanowires (RSW nanowires) at 
experimentally relevant parameter ranges. We derive formulae that determine all 
topological phase boundaries of a multichannel disordered RSW wire. We test 
these formulae with numerical tight-binding simulations at experimentally 
relevant parameter ranges and find good agreement without any fitting 
parameters. We show that there are additional topological transitions for the 
RSW wires leading to a richer phase diagram with further fragmentalization 
beyond that of the \textit{p}-wave models.

\begin{acknowledgments}

%%%%%%%%%%%%%%%%%%%%%%%%%%%%%%%%%%%%%%%%%%%%%%%%
%                                              %
% Acknowledgements                             %
%                                              %
%%%%%%%%%%%%%%%%%%%%%%%%%%%%%%%%%%%%%%%%%%%%%%%%

This work was supported by funds of the Erdal {\.I}n{\"o}n{\"u} chair, by 
T{\"U}B{\.I}TAK under grant No. 110T841, by the Foundation for Fundamental 
Research on Matter (FOM) and by Microsoft Corporation Station Q. \.{I}A is a 
member of the Science Academy---Bilim Akademisi---Turkey; BP, AT and \"{O}B thank 
The Science Academy---Bilim Akademisi---Turkey for the use of their facilities 
throughout this work. 

\end{acknowledgments}

\appendix

%%%%%%%%%%%%%%%%%%%%%%%%%%%%%%%%%%%%%%%%%%%%%%%%
%                                              %
% Appendix                                     %
%                                              %
%%%%%%%%%%%%%%%%%%%%%%%%%%%%%%%%%%%%%%%%%%%%%%%%

\section{Mean free path}\label{SECT:Appendix_MFP_FermiGoldenRule}

%%%%%%%%%%%%%%%%%%%%%%%%%%%%%%%%%%%%%%%%%%%%%%%%
%                                              %
% Appendix Section: Mean free path             %
%                                              %
%%%%%%%%%%%%%%%%%%%%%%%%%%%%%%%%%%%%%%%%%%%%%%%%

\begin{figure}
	\centering
	\includegraphics[width=0.95\columnwidth]{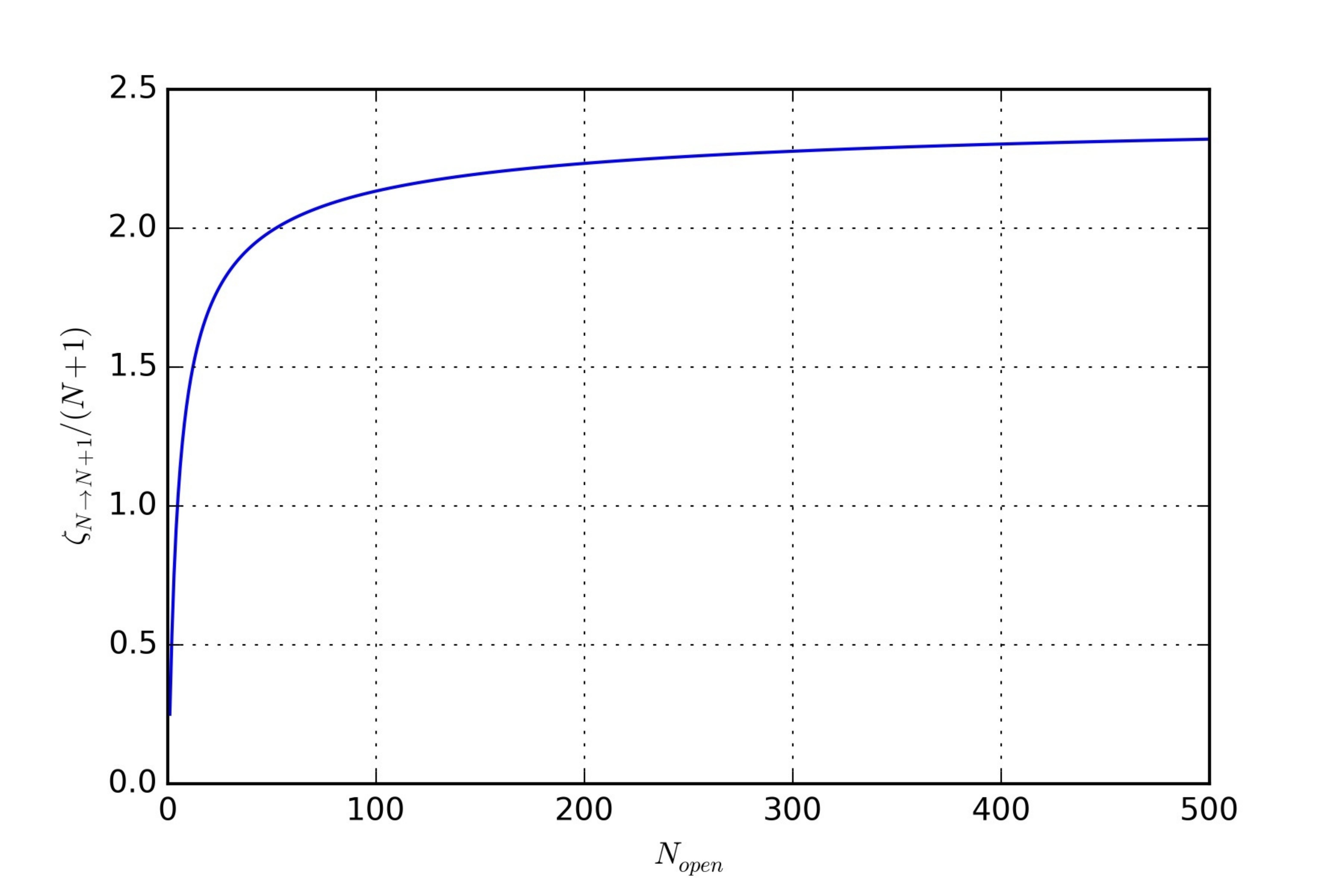}
	\caption{$\zeta_{N\rightarrow N+1}^{-1}/(N+1)$ vs. $N$.}
	\label{FIG:Appendix_Inverse_AlphaperNplusOne}
\end{figure}

We consider a long wire along the $x$-axis, having a length of $L$ along the 
$x$-direction and a width of $W$ along the $y$-direction and metallic leads at 
the end, with a Gaussian disorder of the form $\left\langle 
V(\mathbf{r})\,V(\mathbf{r}')\right\rangle 
=\gamma\,\delta(\mathbf{r}-\mathbf{r}')$. We obtain the ensemble average of the 
matrix element between the $n^\text{th}$ and $l^\text{th}$ transverse channels 
as $\boldsymbol{k}(k_x,n)\rightarrow\boldsymbol{k}'(k_x',l)$ as
\begin{align}
\left\langle |V_{kk'}|^2 \right\rangle&=\frac{\gamma}{LW}\left(1+\frac{\delta_{n,l}}{2}\right).
\end{align}
We then use Fermi's Golden Rule to calculate the inverse lifetime of a momentum 
state $k$, $\tau_{k\rightarrow k'}^{-1}$:
\begin{align}\label{EQN:Appendix:MFP_SingleState}
\left\langle l_{\text{MFP}(k_x,n\rightarrow k_x',l)}^{-1}\right\rangle&=\left(\frac{1}{\hbar}\,\frac{\partial\,\varepsilon_k}{\partial k_x}\right)^{-1}\times\frac{2\pi}{\hbar}\,\frac{\gamma}{LW}\times\nonumber\\ 
 &\qquad \left(1+\frac{\delta_{n,l}}{2}\right)\,\rho(\varepsilon_{k'}).
\end{align}
where $\varepsilon_k$ gives the dispersion relation and $\rho(\varepsilon_{k})$ 
is the density of states. We then sum over the initial and final states $k'$ in 
Eq.~(\ref{EQN:Appendix:MFP_SingleState}) to obtain the total inverse MFP:
\begin{align}\label{EQN:Appendix:TotalInverseMFP_General}
\left\langle l_{\text{MFP}}^{-1} \right\rangle&=\sum_{k_x, k_y; k'_x, k'_y}\left\langle l_{\text{MFP}(k_x,n\rightarrow k_x',l)}^{-1}\right\rangle
\end{align}
We first apply Eq.~(\ref{EQN:Appendix:TotalInverseMFP_General}) to a free 
electron dispersion of the form $\varepsilon(k) = \hbar^2 k^2 / 2m = \hbar^2/2m 
\, (k_{n,x}^2 + n^2 \pi^2 /W^2)$ for $n \in {1, \ldots, \bar{N}}$ where 
$\bar{N}(\mu_\textrm{eff}) = \lfloor W k_F(\varepsilon)/\pi \rfloor$. The 
resulting total ensemble-averaged inverse MFP is
\begin{align}\label{EQN:Appendix:inverseMFP_SumOverChannels_FreeElectron}
\left\langle l_{\text{MFP}}^{-1} \right\rangle	
&=\sum_{n=1}^{\bar{N}}\sum_{l=1}^{\bar{N}} \int\frac{dk'_{n,x}}{\pi/L}\,\frac{m^2}{\hbar^4}\frac{2\gamma W}{L \pi}\,\left(1+\frac{\delta_{nl}}{2}\right)\,\frac{\pi}{W}\times \nonumber\\
&\qquad \frac{\delta(k'_{l,x}\pm\sqrt{2m\varepsilon/\hbar^2-l^2\pi^2/W^2})}{\sqrt{2m\varepsilon/\hbar^2-n^2 \pi^2 /W^2}\,\sqrt{2m\varepsilon/\hbar^2-l^2 \pi^2 /W^2} }\nonumber\\
&= \frac{4m^2\gamma}{\hbar^4\pi k_F}\,\zeta_N^{-1},
\end{align}
where $k_F=\sqrt{2m\varepsilon/\hbar^2}$ is the Fermi wavevector,
\begin{align}\label{EQN:Appendix:inverseZetaN_FreeElectron}
\zeta_N^{-1}&=\frac{3\bar{N}}{2}\sum_{n=1}^{\bar{N}}\eta_n^2+
  2{\bar{N}}\sum_{n=1}^{\bar{N}}\sum_{l>n}^{\bar{N}}
  \eta_n\, \eta_l,
\end{align}
and $\eta_n=\left(\frac{W^2 k_F^2}{\pi^2}-n^2\right)^{-\frac{1}{2}}$, in 
agreement with Eq.(8) in the supporting online material of Rieder \textit{et 
al.}~\cite{REF:Rieder13}. 
The value of $\zeta_{N}$ just below the transition
$N\rightarrow N+1$ (denoted $\zeta_{N\rightarrow 
N+1}$) is plotted in 
Figure~\ref{FIG:Appendix_Inverse_AlphaperNplusOne}. 

We now derive the MFP for a TB dispersion relation given by
\begin{align}\label{EQN:Appendix:TB_Dispersion}
\varepsilon(k_{x,n})&= 2t\,\left(2-\cos{(k_{x,n}a)}-\cos{(n\pi a/W)}\right).
\end{align}

The number of channels is given by $\bar{N}= \lfloor (W/\pi a) 
\arccos{(1-\varepsilon/2t)} \rfloor$ for $0<\varepsilon <4t$ and by 
$\bar{N}= \lfloor (W/\pi a) \arccos{(1-(4-\varepsilon/2t))} \rfloor$ for 
$4t<\varepsilon <8t$. The resulting disorder-averaged inverse MFP reads:
\begin{align}
\left\langle(l_{\text{MFP}}^\text{TB})^{-1}\right\rangle&=\frac{\gamma}{\bar{N}Wa^2t^2}\,(\zeta_N^{\text{TB}})^{-1}
\end{align}
where the dimensionless $(\zeta_N^{\text{TB}})^{-1}$ is given by
\begin{align}\label{EQN:inverseZetaN_TB}
(\zeta_N^{\text{TB}})^{-1}&=\frac{3\bar{N}}{2}\sum_{n=1}^{\bar{N}}(\eta_n^{\text{TB}})^2+\nonumber\\
	&\qquad 2\bar{N}\sum_{n=1}^{\bar{N}}\sum_{l>n}^{\bar{N}} \eta_n^{\text{TB}} \eta_l^{\text{TB}}.
\end{align}
Here, $\eta_n^{\text{TB}}=|\sin{(k_{x,n}\,a)}|^{-1}$ and $\sin{(k_{x,n})}$ is 
obtained using Eq.~(\ref{EQN:Appendix:TB_Dispersion})

\section{Numerical tight-binding simulations}\label{SECT:Appendix_TB}

%%%%%%%%%%%%%%%%%%%%%%%%%%%%%%%%%%%%%%%%%%%%%%%%
%                                              %
% Appendix Section: Numerical tight-binding    %
%                    simulations               %
%                                              %
%%%%%%%%%%%%%%%%%%%%%%%%%%%%%%%%%%%%%%%%%%%%%%%%

We start by obtaining the TB form of the RSW BdG 
Hamiltonian~\cite{REF:deGennes66} in Eq.~(\ref{EQN:SWaveHamiltonian}) in the 
usual way using finite differences (see, for example, 
Ref.\cite{REF:Lutchyn10},\cite{REF:Oreg10},\cite{REF:Stanescu11},\cite{REF:Datta97}). 
It reads:
\begin{align}\label{EQN:Appendix:SWaveHamiltonian_TB}
\mathcal{H}_\mathrm{BdG}^\mathrm{TB} 	&= \left[\left(4t+V(x,y)-\mu(x,y)\right)\,\tau_z + B_\textrm{Z} \, \sigma_z \right. \nonumber\\
										& \quad +\,\left.\Delta(x,y)\, \tau_x \right]\,\left|{x,y}\right\rangle \left\langle{x,y}\right| \nonumber\\
										& \quad +\,\left[ -t\,\tau_z-\frac{i}{2}\alpha_\textrm{SO}(x,y)\,\tau_z\,\sigma_y\right]\,\left|{x+a,y}\right\rangle \left\langle{x,y}\right| \nonumber\\
										& \quad +\,\left[ -t\,\tau_z+\frac{i}{2}\alpha_\textrm{SO}(x,y)\,\tau_z\,\sigma_x\right]\,\left|{x,y+a}\right\rangle \left\langle{x,y}\right| \nonumber\\
										& \quad + \mathrm{h.c.}
\end{align}
where $t = \hbar^2/2 m a^2$ is the hopping parameter, $V(x,y)$ is the Gaussian 
random potential, $\mu(x,y)$ is the relevant gate potential, $B_\textrm{Z}$ is 
the Zeeman field, $\Delta(x,y)$ is the \textit{s}-wave superconducting pairing 
(taken to be real), $\alpha_\textrm{SO}(x,y)$ is the effective Rashba SOC due 
to proximity effect and $a$ is the lattice constant for the TB lattice. Here, 
$V(x,y)$, $B_\textrm{Z}$, $\Delta(x,y)$ and $\alpha_\textrm{SO}(x,y)$ are 
nonzero only within the scattering region. $B_\textrm{Z}$, 
$\alpha_\textrm{SO}(x,y)$ and $\Delta(x,y)$ are constant within the scattering 
region except for the values of $\alpha_\textrm{SO}(x,y)$ in the scattering 
region-lead boundary, where we take it to be half of its value in the bulk. 

The experimental values for InSb nanowires quoted in  Mourik \textit{et 
al.}~\cite{REF:Mourik12} are $\alpha_\textrm{SO}=0.2\,\textrm{eV}\textup{\AA}$, 
$l_\textrm{SO}\sim 2000\textup{\AA}$, $\Delta = 0.25 m\textrm{eV}$, $E_Z/B = 
1.5 m\textrm{eV}/\textrm{T}$, $m_* = 0.015m_e$ and $\alpha_\textrm{SO}^2 m_* / 
2 \hbar^2 \sim 0.04 m\textrm{eV}$. We employ these values verbatim, except for 
$l_\textrm{SO}$ (and correspondingly, $\alpha_\textrm{SO}$), for which we use 
parameters much more accessible experimentally.

We use the Kwant library~\cite{REF:Kwant14} to obtain the topological phase 
diagram in our numerical plots. The Kwant library can extract the scattering 
matrix (\textit{S}-matrix),\cite{REF:Datta97} and therefore the reflection 
matrix (\textit{r}-matrix) for a given tight-binding system with leads. The 
topological index $Q_\textrm{D}$ can be obtained from the \textit{r}-matrix 
through $Q_\textrm{D}=\textrm{det}(r)$ (see Ref.~\cite{REF:Fulga11}).

The numerical parameters quoted in the caption of 
Fig.~\ref{FIG:SWave_TB_NumericalAnalytical_MuvsB_Zoomed} correspond to 
$t=1.5m\textrm{eV}$, $a=40.8\textrm{nm}$, $l_\textrm{SO}=4.08\mu\textrm{m}$ and 
$\alpha=6.3\times 10^{-6}\,c$. Disregarding screening, a Zeeman Energy of, say, 
$0.35t$ on the plot would correspond to a magnetic field $0.35\textrm{T}$, a 
value easily accessible by the experiment. In 
Figures~\ref{FIG:SWave_TB_Analytical_Numerical_Combined}, 
\ref{FIG:Appendix:SWave_MuvsV0_Analytical_VaryingB}, 
\ref{FIG:Appendix:SWave_TB_Analytical_MuvsB_FourChannels} and 
\ref{FIG:Appendix:SWave_TB_NumericalAnalytical_MuvsB_FullBandwidth}	, 
$l_\textrm{SO}=6.0\mu\textrm{m}$, $t=0.7m\textrm{eV}$, $a=60.0n\textrm{m}$ and 
$\alpha=4.2\times 10^{-6}\,c$. A Zeeman energy of $0.35t$ corresponds to 
$B=0.17\,\textrm{T}$.

The TB form of the effective PW Hamiltonian of 
Eq.~(\ref{EQN:Appendix:PWaveHamiltonian}) used in 
Appendix~\ref{SECT:Appendix_PWave} is as follows:
\begin{align}\label{EQN:Appendix:PWaveHamiltonian_TB}
\mathcal{H}_\mathrm{PW}^\mathrm{TB}	&= \left[4t+V(x,y)-\mu(x,y) \right]\,\tau_z\,\left|{x,y}\right\rangle \left\langle{x,y}\right| \nonumber\\
									& \quad +\,\left[ -t\,\tau_z-\frac{i}{2}\Delta_\textrm{eff}(x,y)\,\tau_x\right]\,\left|{x+a,y}\right\rangle \left\langle{x,y}\right| \nonumber\\
									& \quad +\,\left[ -t\,\tau_z-\frac{i}{2}\Delta_\textrm{eff}(x,y)\,\tau_y\right]\,\left|{x,y+a}\right\rangle \left\langle{x,y}\right| \nonumber\\
									& \quad + \mathrm{h.c.}
\end{align}
We use numerical values similar to the RSW case in our PW simulations, except 
to impose $\Delta_\textrm{eff} = \Delta \, \alpha_\textrm{SO} / 
\sqrt{B^2-\Delta^2}$.

\section{Topological phase diagram over the full bandwidth}\label{SECT:Appendix_FullBWPlots}

%%%%%%%%%%%%%%%%%%%%%%%%%%%%%%%%%%%%%%%%%%%%%%%%
%                                              %
% Appendix Section: Topological phase diagram  %
%                    over the full bandwidth   %
%                                              %
%%%%%%%%%%%%%%%%%%%%%%%%%%%%%%%%%%%%%%%%%%%%%%%%

In this section, we present plots of the topological phase diagram that we 
obtain analytically from Eq.~(\ref{EQN:Q_D}) using a TB dispersion relation (see 
Section~\ref{SECT:DisorderedTSWires}) over the full bandwidth. Although only 
the low $\mu$ regions in our plots correspond to experimentally relevant 
nanowires, the full bandwidth range would be important for systems that are 
inherently TB, such as atomic chains~\cite{REF:Nadj-Perge14} or photonic 
metamaterials~\cite{REF:Tan14} simulating topological 
properties.~\cite{REF:Lu14} All analytical plots are produced using 
Eq.~(\ref{EQN:Q_D}) (Eq.~(\ref{EQN:Appendix:QChiral_PWave_Disordered}) for the PW 
case), but using a TB dispersion relation for $\epsilon(p)$ in the relevant 
expressions. All of the numerical results are obtained using a TB simulation 
utilizing Kwant software, as discussed in the main text.
 
\begin{figure}
	\includegraphics[width=1.0\columnwidth]{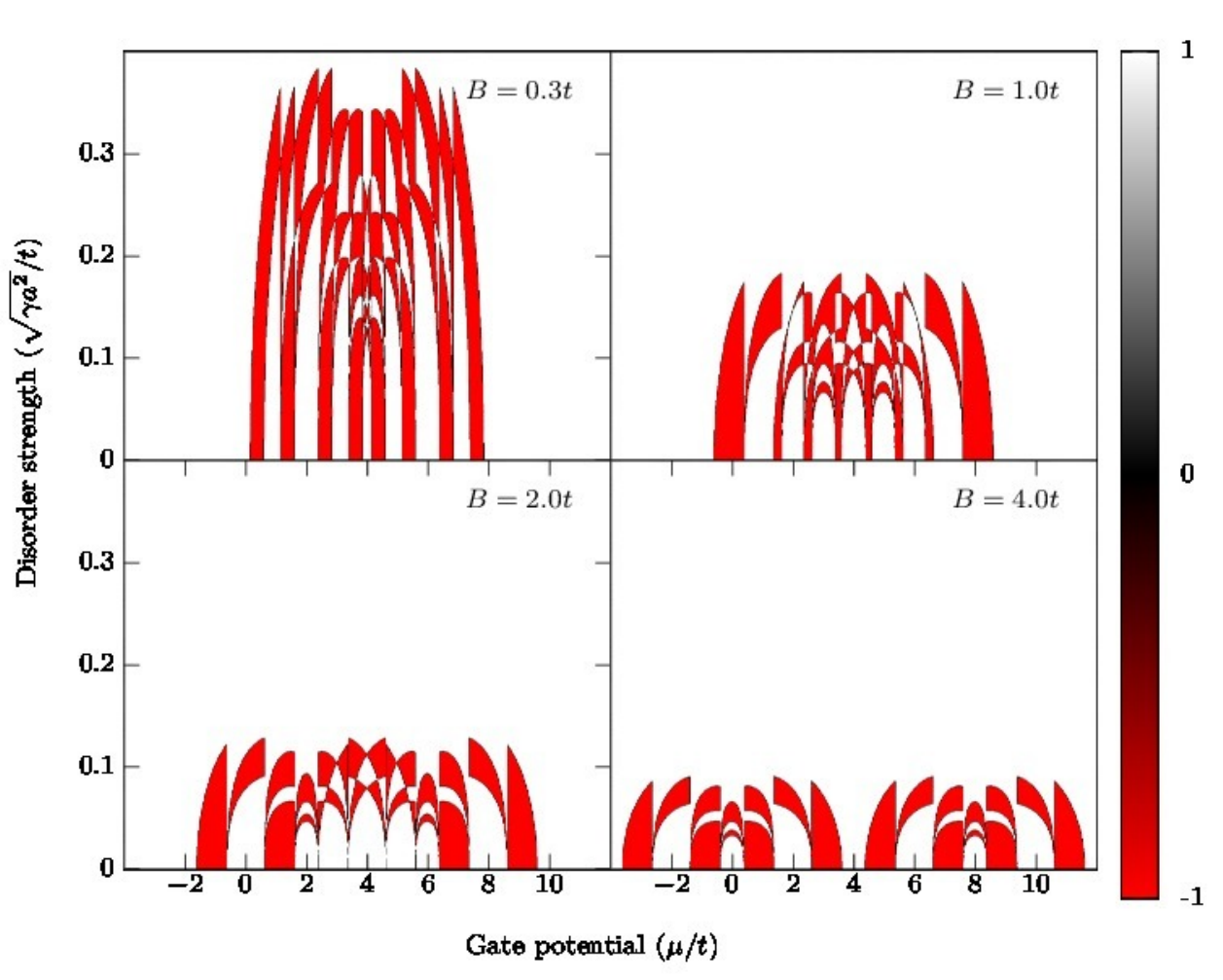}	
	\caption{(Color online) $\mu$ vs. $V_0 = \sqrt{\gamma/a^2}$ vs. 
	$Q_\textrm{D}$ for a multichanneled RSW wire for different $B$, obtained 
	analytically using Eq.~(\ref{EQN:Q_D}). a), b) Low 
	magnetic field $B\gtrsim\Delta$ limit requires a full RSW model and 
	topological order can survive up to high disorder strengths. c), d) The 
	spin-polarized system can be described by a PW model and topological order 
	is completely destroyed with less disorder. Here, $W=4a$, 
	$\alpha_\textrm{SO}=0.015\hbar/ma$ and $\Delta = 0.20t$ where $t=\hbar^2/2ma^2$ and 
	$a$ is the tight-binding lattice spacing. See Appendix~\ref{SECT:Appendix_TB} 
	for a discussion of corresponding experimental parameters.
	}\label{FIG:Appendix:SWave_MuvsV0_Analytical_VaryingB}
\end{figure}

\begin{figure}
	\includegraphics[width=1.0\columnwidth]{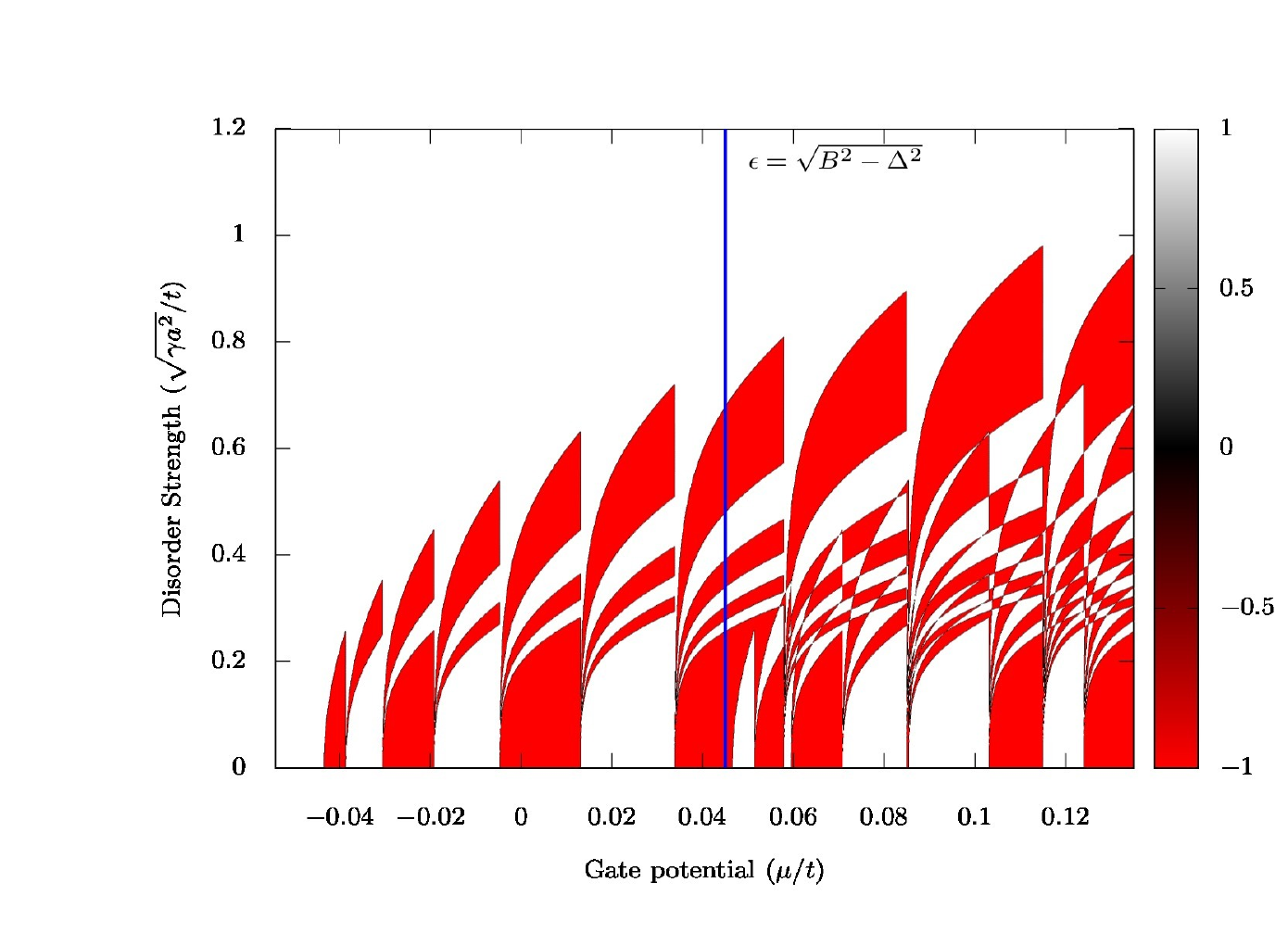}	
	\caption{(Color online) $\mu$ vs. $V_0 = \sqrt{\gamma/a^2}$ vs. 
	$Q_\textrm{D}$ for a multichanneled wide RSW wire, obtained obtained 
	analytically using Eq.~(\ref{EQN:Q_D}), with $W=77a$.
	Here, $\alpha_\textrm{SO}=0.015\hbar/ma$ ($l_\textrm{SO}=100a$), 
	$\Delta = 0.20t$ and $B=0.205t$ with the hopping parameter 
	$t=\hbar^2/2ma^2=0.7$ $m$eV and the lattice spacing $a=60$ $n$m. The 
	blue vertical line at $\mu = \epsilon = \sqrt{B^2-\Delta^2}$ is the 
	bottom of the second spin band.
	}\label{FIG:Appendix:SWave_MuvsV0_Analytical_LargeWidth}
\end{figure}

Figure~\ref{FIG:Appendix:SWave_MuvsV0_Analytical_VaryingB} depicts the 
analytically calculated topological phase diagram for an RSW wire as a function 
of $\mu$ and the disorder strength, for various magnetic field strengths. The 
transition between a RSW wire and a pair of oppositely polarized PW wires can 
be seen as increasing magnetic field polarizes the system. The topological 
order is less robust against disorder for higher magnetic fields, because the 
coherence length becomes longer with increasing $B$. This is the reason why the 
spin polarized regimes where PW model applies is typically less robust than the 
lower field regimes where both spin species exist as seen in 
Fig.~\ref{FIG:Appendix:SWave_MuvsV0_Analytical_VaryingB}(a) and 
\ref{FIG:Appendix:SWave_MuvsV0_Analytical_VaryingB}(c) or (d). In order to 
complete the discussion, we also present an analytical plot 
(Figure~\ref{FIG:Appendix:SWave_MuvsV0_Analytical_LargeWidth}) for an RSW wire 
for which B is greater than the subband spacing but less than the bandwidth. 
While this regime is experimentally very hard to achieve, it is useful for 
comparing the PW and the RSW regimes. The vertical blue line denotes the bottom 
of the higher energy spin band beyond which both spin species exist. We note 
that the critical disorder strength increases with the chemical potential, 
hence spin-polarized regime, which appear at lower chemical potential values, 
is less robust against disorder.

\begin{figure}
	\centering 
	\includegraphics[width=\columnwidth]{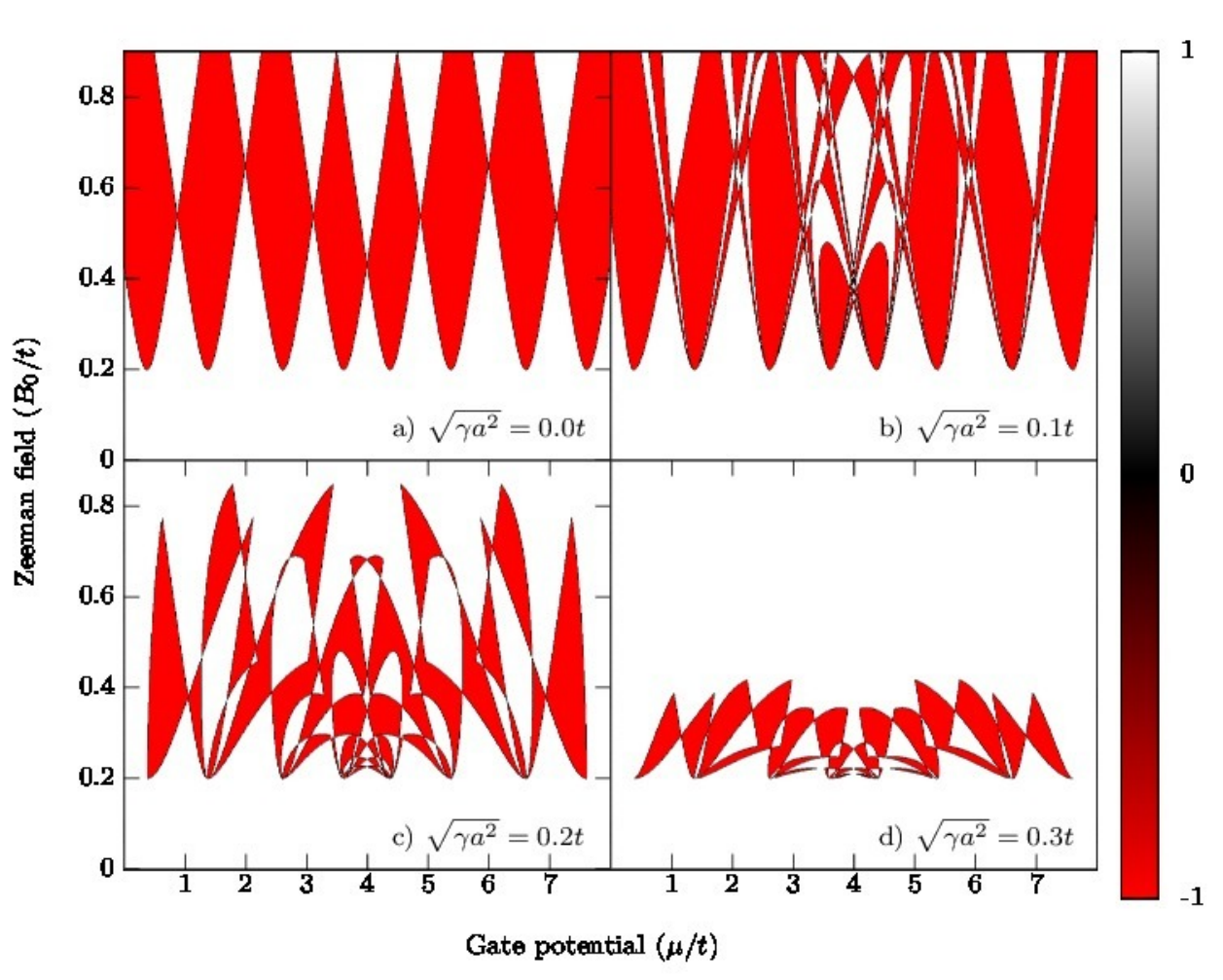} 	
	\caption{(Color online) $\mu$ vs. $B$ vs. $Q_\textrm{D}$ for varying disorder 
	strengths for an RSW TS with Gaussian disorder, 
	analytically calculated using Eq.~(\ref{EQN:Q_D}) for a 
	four-channel TB system. Subfigure c) matches the numerical data shown in 
	Fig.~\ref{FIG:Appendix:SWave_TB_NumericalAnalytical_MuvsB_FullBandwidth}	. The 
	parameters used are $\alpha_\textrm{SO}=0.015\hbar/ma$ and $\Delta=0.2t$, where $t=\hbar^2/2ma^2$
	$a$ is the lattice spacing. See Appendix~\ref{SECT:Appendix_TB} for a 
	discussion of corresponding experimental parameters.
	}\label{FIG:Appendix:SWave_TB_Analytical_MuvsB_FourChannels}
\end{figure}

In Figure~\ref{FIG:Appendix:SWave_TB_Analytical_MuvsB_FourChannels}, the 
analytically calculated phase diagram of a wire with $W=4a$ is plotted with 
increasing disorder. We see that the phase diagram gets fragmented as number of 
channels are increased. We also note that for a given amount of disorder, there 
is a maximum Zeeman field $B_\text{max}$ above which no topological order is 
present. The reason is that in our numerical TB simulations, the localization 
length is not a monotonous function of energy. It grows (with increasing 
energy) until the middle of the band, and after that it decreases as the energy 
comes closer to the band edge. This places an upper magnetic field limit to 
topological regions since the superconducting coherence length monotonically 
increases with $B$. For a pure quadratic dispersion, the upper limit is given 
by the limitations of the approximations of Fermi's Golden Rule and would 
increase indefinitely with increasing energy as discussed in the main text. We 
note that the upper limit discussed here has a different origin than that 
discussed by Ref.\cite{REF:Rainis13} for finite-length wires. 

\begin{figure}
	\centering
	\hspace*{-1cm} 
	\includegraphics[width = 0.98\columnwidth]{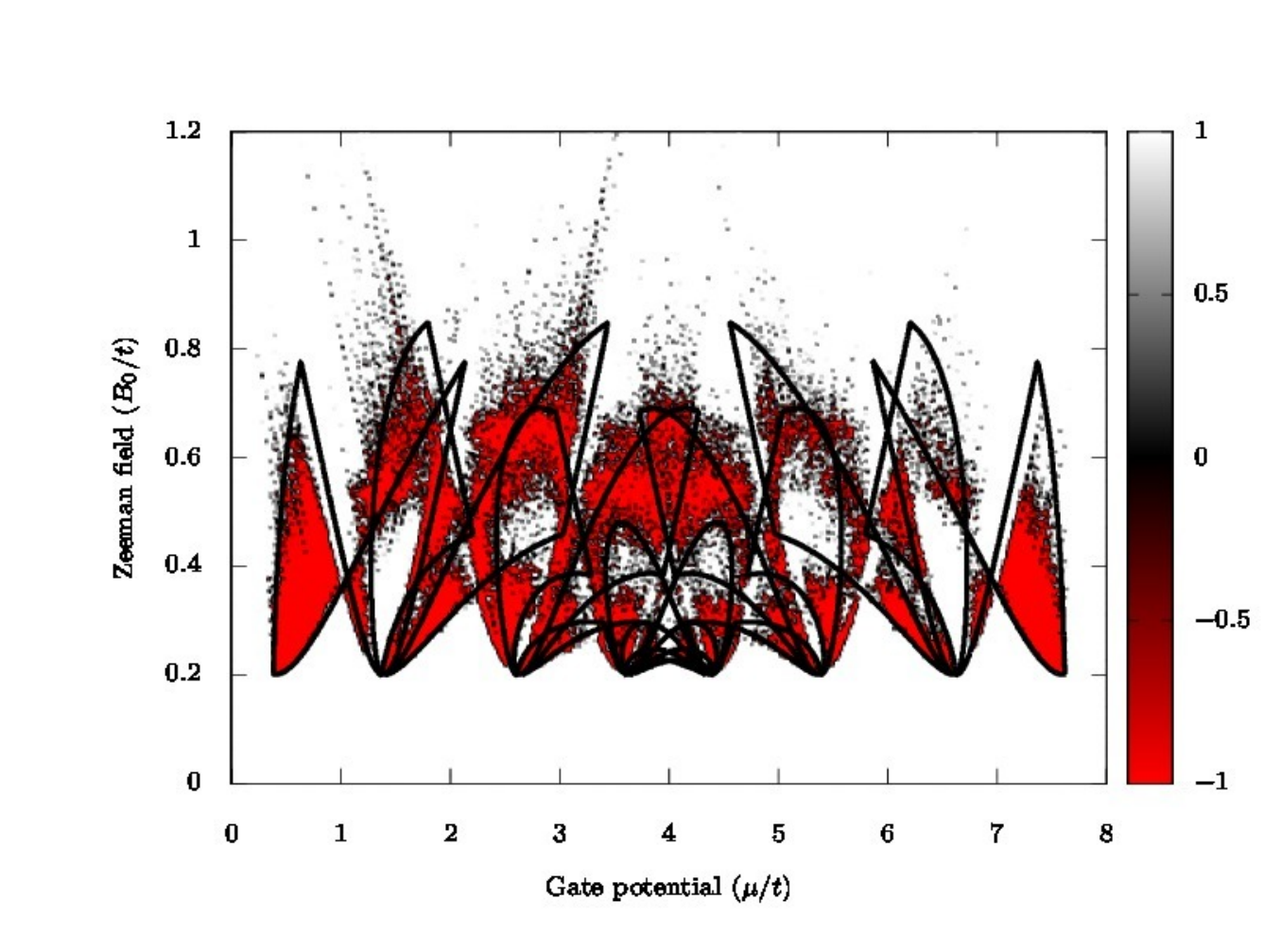}	
	\caption{(Color online) $\mu$ vs. $B$ vs. $Q_\textrm{D}$ for a four-channel 
	system (compare with Figs.~\ref{FIG:SWave_TB_NumericalAnalytical_MuvsB_Zoomed} 
	and \ref{FIG:Appendix:SWave_TB_Analytical_MuvsB_FourChannels}). The black lines, 
	which represent topological boundaries, are obtained analytically using 
	Eq.~(\ref{EQN:Q_D}). The background red-white colors are obtained using 
	tight-binding numerical simulations. The parameters are $V_0=0.2t$, 
	$\Delta = 0.2t$ and $\alpha_\textrm{SO}=0.015\hbar/ma$. See 
	Appendix~\ref{SECT:Appendix_TB} for a discussion of corresponding 
	experimental parameters.
	}\label{FIG:Appendix:SWave_TB_NumericalAnalytical_MuvsB_FullBandwidth}		
\end{figure}

We finally present the full TB bandwidth version of 
Fig.~\ref{FIG:SWave_TB_NumericalAnalytical_MuvsB_Zoomed}, with slightly 
different material properties, here in 
Fig.~\ref{FIG:Appendix:SWave_TB_NumericalAnalytical_MuvsB_FullBandwidth}. This 
figure is the numerical simulation result that matches the last of the 
analytical plots in 
Fig.~\ref{FIG:Appendix:SWave_TB_Analytical_MuvsB_FourChannels}. The relevant 
numerical values are given in each of the Figures' captions.

\section{Topological phase diagram for multichannel effective \textit{p}-wave nanowires with disorder}\label{SECT:Appendix_PWave}

%%%%%%%%%%%%%%%%%%%%%%%%%%%%%%%%%%%%%%%%%%%%%%%%
%                                              %
% Appendix Section: Disordered PWave           %
%                                              %
%%%%%%%%%%%%%%%%%%%%%%%%%%%%%%%%%%%%%%%%%%%%%%%%

\begin{figure}
	\centering 
	\hspace*{-1.5cm} 
	\includegraphics[width=1.0\columnwidth]{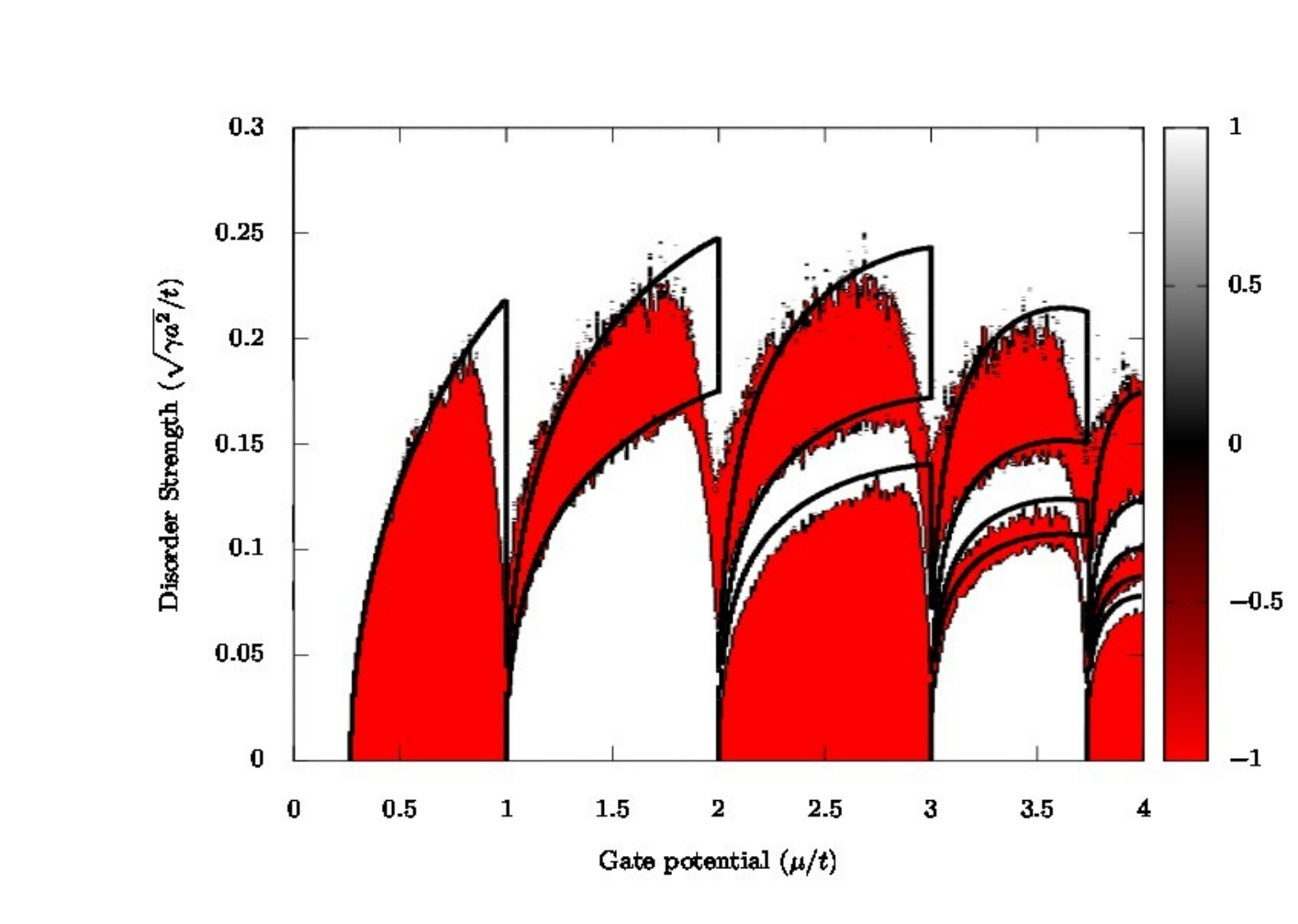} 
	\caption{(Color online) $\mu$ vs. $\sqrt{\gamma/a^2}$ vs. $Q$ for a 
	multichanneled PW wire with dimensions $W=4a$ and $L=60000a$ ($L$ used only 
	in the numerical tight-binding code) and with 
	$\alpha_\textrm{SO}=0.01\hbar/ma$, where $a$ is the tight-binding lattice 
	spacing. The red-white colors in the background are obtained numerically 
	with a tight-binding method whereas the black solid lines are obtained 
	using Eq.~(\ref{EQN:Appendix:QChiral_PWave_Disordered}) with 
	Eq.~(\ref{EQN:inverseMFP_TB_EndResult}).}
	\label{FIG:Appendix:PWave_TB_Analytical_Numerical_Combined}
\end{figure}

In this Appendix section, we present the effects of disorder on PW wires, which 
is a system previously studied in literature,~\cite{REF:Adagideli14, 
REF:Akhmerov11,REF:Brouwer11b,REF:DeGottardi13a,REF:Fulga11,REF:Hui14a, 
REF:Lobos12,REF:Rieder13,REF:Potter11a,REF:Potter11b,REF:Rieder12,REF:Sau12, 
REF:Sau13} for completeness and for comparison with the results of our paper 
for disordered multichannel RSW nanowires. We start with the Hamiltonian in 
Eq.~(\ref{EQN:Appendix:PWaveHamiltonian}) and present the topological charge in 
Eq.~(\ref{EQN:Appendix:QChiral_PWave_Disordered}). We plot the topological phase 
diagram for a PW wire as a function of $\mu$ and disorder strength for a fixed 
$B_\textrm{Zeeman}$ 
(Fig.~\ref{FIG:Appendix:PWave_TB_Analytical_Numerical_Combined}) and compare 
this plot with its analogue for RSW wires 
(Fig.~\ref{FIG:SWave_TB_Analytical_Numerical_Combined}).

The BdG Hamiltonian for an effective \textit{p}-wave wire with spatially 
homogeneous effective SOC strength is
\begin{align}\label{EQN:Appendix:PWaveHamiltonian}
\mathcal{H}_\mathrm{BdG}^\mathrm{PW} &=\varepsilon(p)\,\tau_z+\Delta_\textrm{eff}\,\mathbf{p}\cdot\mathbf{\tau}.
\end{align}
Note that $\Delta_\textrm{eff}$ has units of velocity while $\Delta$ in 
Eq.~(\ref{EQN:SWaveHamiltonian}) has units of energy. This effective SOC 
strength is related to the corresponding RSW superconducting gap by 
$\Delta_\textrm{eff} = \Delta \, \alpha_\textrm{SO} / 
\sqrt{B^2-\Delta^2}$.~\cite{REF:Lutchyn10} We consider a Gaussian disorder of 
the form $\left\langle V(\mathbf{r}) \, V(\mathbf{r}') \right\rangle = 
\gamma\,\delta(\mathbf{r}-\mathbf{r}')$ for $\mathbf{r},\mathbf{r}'$ in the 
wire, with $\gamma$ as the disorder strength and $\left\langle 
V(\mathbf{r})\right\rangle=0$. This Hamiltonian is useful for comparison with 
the fully polarized limit of the RSW case.

The Hamiltonian in Eq.~(\ref{EQN:Appendix:PWaveHamiltonian}) is in 
Altland-Zirnbauer (AZ) symmetry class D in two dimensions~\cite{REF:Altland97} 
with a $\mathbb{Z}_2$ topological number. This Hamiltonian also possesses a 
chiral symmetry, broken by the $\Delta_\textrm{eff}\, p_y \tau_y$ term. If this 
term is set to zero, the Hamiltonian is also in class 
BDI~\cite{REF:Rieder12,REF:Tewari12,REF:Diez12,REF:Rieder13} having a 
$\mathbb{Z}$ topological number. (1D wires trivially satisfy this condition.) 
In the thin wire limit, i.e. $\Delta_\textrm{eff}\ll\hbar/mW$, the chiral 
symmetry breaking term is $\mathcal{O}\left((m\Delta_\textrm{eff} 
W/\hbar)^2\right)$. The wire in class BDI can have an integer number of 
Majorana fermions at its ends. The chiral symmetry breaking term pairwise 
hybridizes these solutions. Hence the chiral topological number 
$Q_\textrm{BDI}\in \mathbb{Z}$ and the class-D topological number 
$Q_\textrm{D}\in \mathbb{Z}_2$ are related as $Q_\textrm{D} = 
{-1}^{Q_\textrm{BDI}}$.~\cite{REF:Fulga11}

In order to solve the Schr\"{o}dinger equation $H\,\Psi = E\,\Psi$ at $E = 0$ 
to obtain the Lyapunov exponents, we follow Adagideli \textit{et 
al.}~\cite{REF:Adagideli14} to off-diagonalize the Hamiltonian and apply an 
imaginary gauge transformation. This allows us to re-express $Q_\textrm{BDI}$ 
in terms of $\Lambda_n$:~\cite{REF:Rieder13}
\begin{align}\label{EQN:Appendix:QChiral_PWave_Disordered} 
Q_\textrm{BDI}&=\sum_{n=1}^{\bar{N}}\Theta\left(\xi-\frac{1}{\Lambda_n}\right), 
\end{align}
where $\bar{N}=\lfloor W/\pi\sqrt{2m\mu/\hbar^2}\rfloor$ and $\lfloor x 
\rfloor$ is the usual floor function. We obtain $\Lambda_n$ again using 
Eq.~(\ref{EQN:LyapunovExponent_from_MFP}). We obtain $l_\text{MFP}^{-1}$ using 
Fermi's Golden Rule (see Appendix~\ref{SECT:Appendix_MFP_FermiGoldenRule}) 
first for a quadratic dispersion relation and then for a TB dispersion 
relation. 

We compare the results found using Eq. 
\ref{EQN:Appendix:QChiral_PWave_Disordered} with those obtained by numerical 
simulations in Figure \ref{FIG:Appendix:PWave_TB_Analytical_Numerical_Combined} 
and find an excellent fit over the whole TB bandwidth. In a clean PW wire 
($\sqrt{\gamma/a^2}=0$), Majorana modes appear if $\bar{N}$ is odd and Majorana 
states fuse to form ordinary Dirac fermions if $\bar{N}$ is even. This behavior 
survives up to a finite disorder strength (see 
Fig.~\ref{FIG:Appendix:PWave_TB_Analytical_Numerical_Combined}). As in the case 
of RSW wires, further increase of the disorder strength gives a series of 
transitions between non-trivial and trivial topological phases as each 
$\Lambda_n$ increases and crosses $\xi^{-1}$. 
While both multichanneled RSW and PW wires
feature reentrant behavior, we see that there are additional transitions for the
RSW wires leading to a richer phase diagram (compare 
Figures~\ref{FIG:Appendix:PWave_TB_Analytical_Numerical_Combined} 
and~\ref{FIG:SWave_TB_Analytical_Numerical_Combined}), in agreement with our
analytical results presented in Eq.~(\ref{EQN:Topological_phase_boundaries_TB}).

\end{document}